\documentclass{iau-JDSS}
\usepackage{graphicx}

\title[JD 4.~~CP and related phenomena in the context of Stellar Evolution]{CP and related phenomena in the context of Stellar Evolution}

\author[\,\,\,\,\,\,\,\,\,\,\,\,\,\,\,\,\,\,\,\,\,Braithwaite, Akg\"{u}n, E.~Alecian, Kholtygin, Mathis, Michaud, Landstreet et al.]
{J.~Braithwaite$^{1,2}$,
T.~Akg\"{u}n$^3$,
E.~Alecian$^4$,
A.F.~Kholtygin$^5$, 
J.D.~Landstreet$^{6, 7}$,
S.~Mathis$^8$,
G.~Michaud$^9$,
J.~Portnoy$^{10}$,
G.~Alecian$^4$,
V.D.~Bychkov$^{11}$,  
L.V.~Bychkova$^{11}$
N.~Drake$^5$, 
S.N.~Fabrika$^{11}$, 
A.~Reisenegger$^3$, 
R.~Steinitz$^{12}$,
M.~Vick$^9$
}

\affiliation{
$^1$ Canadian Institute for Theoretical Astrophysics, Toronto, Canada \\[\affilskip]
$^2$ Argelander Institut f\"ur Astronomie, Bonn, Germany; {\tt jonathan@astro.uni-bonn.de} \\[\affilskip]
$^3$ Departamento de Astronom\'{i}a y Astrof\'{i}sica, Pontificia Universidad Cat\'{o}lica de Chile, Santiago, Chile; {\tt akgun@astro.puc.cl} \\[\affilskip]
$^4$Observatoire de Paris, LESIA, 5 place Jules Janssen, 92190 Meudon, France {\tt evelyne.alecian@obspm.fr} \\[\affilskip]
$^5$Department of Astronomy, Saint-Petersburg State University, Saint-Petersburg, Russia; {\tt afkholtygin@gmail.com} \\[\affilskip]
$^6$University of Western Ontario, 1151 Richmond Street, London ON N6A 3K7, Canada \\[\affilskip]
$^7$Armagh Observatory, Northern Ireland \\[\affilskip]
$^8$Laboratoire AIM, CEA/DSM-CNRS-Universit\'e Paris Diderot, IRFU/SAp Centre de Saclay, F-91191 Gif-sur-Yvette, France; {\tt stephane.mathis@cea.fr} \\[\affilskip]
$^9$D\'epartement de Physique, Universit\'e de Montr\'eal, Montr\'eal, PQ, H3C~3J7, Canada; {\tt michaudg@astro.umontreal.ca} \\[\affilskip]
$^{10}$Sami Shamoon College of Engineering, Israel {\tt jacovp@sce.ac.il} \\[\affilskip]
$^{11}$Special Astrophysical Observatory, Russia \\[\affilskip]
$^{12}$Physics Dept., Ben Gurion University of the Negev, Israel}
 
\pubyear{2009}
\volume{Volume ?}  
\pagerange{1-10}
\date{28 September 2009}
\jname{Highlights of Astronomy, Volume ?}
\editors{?, eds.}
\begin{document}

\maketitle

\begin{abstract}
We review the interaction in intermediate and high mass stars between their evolution and magnetic and chemical properties.
We describe the theory of Ap-star `fossil' fields, before touching on the expected secular diffusive processes which give rise to evolution of the field. We then present recent results from a spectropolarimetric survey of Herbig Ae/Be stars, showing that magnetic fields of the kind seen on the main-sequence already exist during the pre-main sequence phase, in agreement with fossil field theory, and that the origin of the slow rotation of Ap/Bp stars also lies early in the pre-main sequence evolution; we also present results confirming a lack of stars with fields below a few hundred gauss. We then seek  which macroscopic motions compete with atomic diffusion in determining the surface abundances of AmFm stars. While turbulent transport and mass loss, in competition with atomic diffusion, are both able to explain observed surface abundances,  the interior abundance distribution is different enough to potentially lead to a test using asterosismology. Finally we review progress on the turbulence-driving and mixing processes in stellar radiative zones.
\keywords{stars: magnetic fields -- stars: chemically peculiar -- stars: evolution -- ({\it magnetohydrodynamics}) MHD --
stars: rotation -- stars: intermediate mass -- stars: pre-main sequence}
\end{abstract}

\firstsection 

\section{The interesting lives of middle main sequence stars}

Observations reveal that around 5-10\% of intermediate-mass main-sequence (MS) stars (1.6 - 8$\,M_\odot$) have magnetic fields of strength 300 G - 30 kG (see Donati \& Landstreet 2009 for a recent review of the so-called `chemically peculiar' (CP), or Ap and Bp stars). The fields are large-scale and apparently static, i.e. they do not evolve at all over periods of decades, and their host stars rotate much more slowly than normal A stars. 

It is worthwhile to look at magnetic middle MS stars in the broader context of stellar evolution. Stars form by gravitational collapse on a timescale of a few 0.1 Myr, set by the free-fall time. At a size of a few 100 AU, angular momentum leads to slower growth through quasi-static disk accretion. The proto-star shrinks towards the MS as a T Tau star, and later, if massive enough, as a Herbig AeBe star. During early stages of collapse much magnetic flux may be retained, or lost via ambipolar diffusion, but it is not clear how to retain flux whilst the star is almost fully convective, or how the field of some Herbig AeBe stars might be related to the earlier T Tau dynamo activity phase.

On the MS, surface chemistry changes occur as a result of gravitational and radiatively driven atomic diffusion, which must compete with convection, meridional circulation and turbulent diffusion, mass loss from the surface, the magnetic field (especially in the upper atmosphere and beyond), and possibly even accretion. The complications of this multi-dimensional competition probably explain why some stars with fossil fields are marked by very distinctive chemical peculiarities (hence `CP' stars) while other, hotter magnetic stars are not, but the details of the effects are still far from clear. However, the way in which different effects predominate with changes in stellar parameters such as mass, age and rotation rate make these stars excellent laboratories for the study of the various important internal physical processes.

Magnetic fields have been known in peculiar B stars up to about 8 $M_\odot$ for decades, but are now being found in a (still small) fraction of stars well above this mass, mostly without chemical peculiarities (e.g. Bouret et al. 2008 and refs. therein). These fields have significant effects on the strong stellar winds, and in extreme cases wind gas is trapped in closed field loops (Landstreet \& Borra 1978; Babel \& Montmerle 1997); this effect has been beautifully modelled in detail by Townsend et al. (2007).

The host stars of these fossil fields straddle the boundary between low mass stars which have greatly increased luminosity as giants, and high-mass stars that evolve at almost constant luminosity as supergiants. Note, that during the MS phase, R increases by a factor of 2, and $T_{\rm eff}$ decreases by 30\%. The magnetic fields also pose major puzzles. Why are no fields, dynamo or fossil, found in MS F3 - F5 stars? Why do fields in MS Ap stars appear to decay on a time scale short compared to the Cowling time but still a significant fraction of the MS lifetime (Landstreet et al. 2008)? How and when do strong magnetic fields influence rotation rates? What happens to fields as the host stars evolve to the giant phase -- can the fossil field anchor to stable layers deep in the star and persist in spite of the deep convection? Can fossil fields survive the giant phases to the end, to become the fossil fields of white dwarfs, or do these have a different origin, perhaps in a close binary system?

\section{Fossil fields in Ap stars and degenerate stars}\label{sec:equilibria}

Over the 60 years since magnetic fields were discovered in Ap stars the subject has accompanied the development of magnetohydrodynamics, as theoretical explanations have been sought to explain the properties of the observed `oblique rotators' -- large-scale, static fields. Two theories have been proposed. In {\it core-dynamo theory} the field is generated in the core by differential rotation and convection, as is known to work in the solar envelope, and rises through the radiative zone to the surface -- and herein lies a severe drawback (another being the lack of positive correlation between field strength and rotation speed): There are diffusive mechanisms to cause this rise, but the timescales are apparently too long. According to {\it fossil theory}, what we see is a remnant from either the original molecular cloud or a protostellar dynamo -- when the star settles down, the field relaxes into a stable equilibrium, which then evolves only via the same very slow diffusive processes over timescales longer than the MS lifetime.

\begin{figure}
\parbox[t]{0.31\hsize}{\caption{\small Field lines of two magnetohydrostatic equilibria. The equilibrium on the left is axisymmetric, and can exist in a star (giving a dipolar field on the surface) as well as inside a non-gravitating bubble. On the right, a more complex non-axisymmetric equilibrium, which can only exist in a star. [Blue shading represents surface of star.]
\label{fig:jon}}}\,\,\,\,\,
\parbox[t]{0.65\hsize}{\mbox{}\\
\includegraphics[width=0.5\hsize,angle=0]{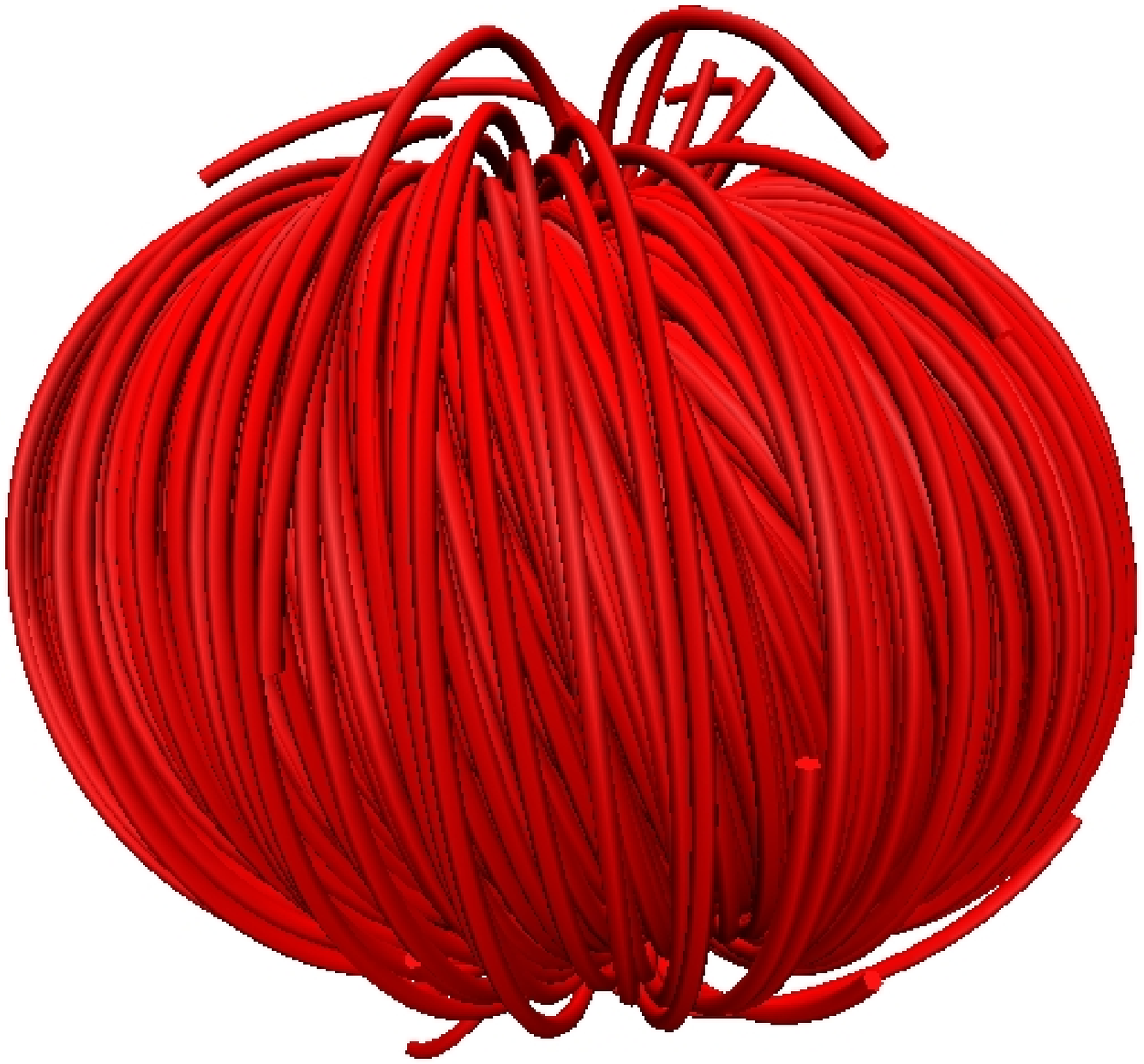}
\includegraphics[width=0.5 \hsize,angle=0]{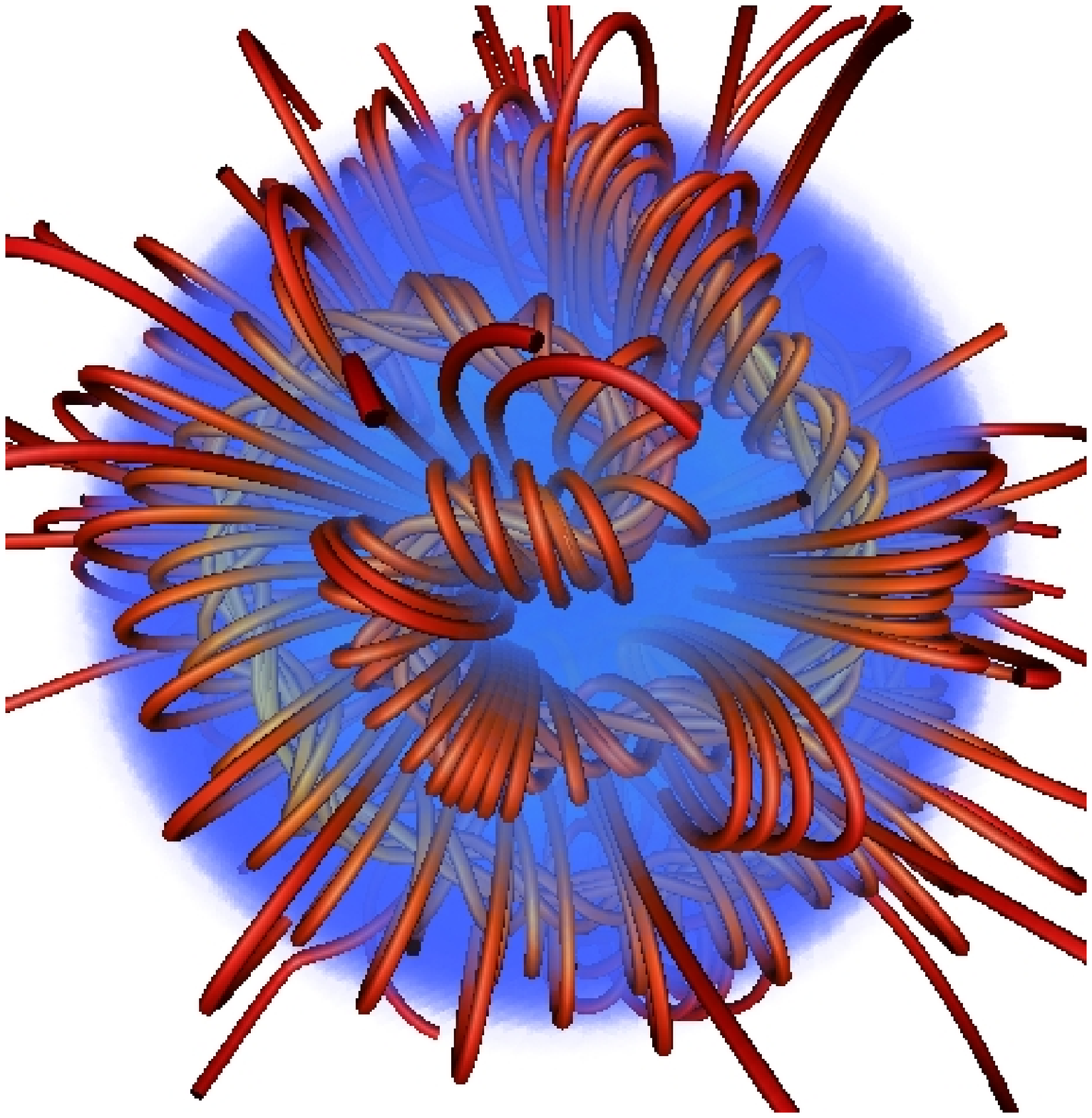}}
\end{figure}

Theoretical work on fossil fields has focussed on demonstrating that such stable equilibria are in fact possible. This is problematic, as it can be shown that the simplest magnetic field configurations are always unstable (Goossens et al. 1981). Generally studies have looked only at axisymmetric fields: Tayler (1973) showed that all purely toroidal fields are prone to interchange (axisymmetric) and kink (non-axisymmetric) instabilities, and Markey \& Tayler (1973) and Wright (1973) showed that purely poloidal fields are also unstable. Therefore it was concluded that a stable axisymmetric field (if it exists) must indeed have both poloidal and toroidal components. Now, such a field can be expressed as the sum of a poloidal and a toroidal component, each completely described by a single scalar function. The Lorentz force for an axisymmetric field cannot have an azimuthal ($\phi$) component, since no counterpart exists in the hydrostatic fluid forces that can balance it. From this it can be shown that the two functions defining the poloidal and toroidal parts must be related. One consequence of this is that the toroidal field is entirely confined within the poloidal field lines that close inside the star. Simulations have now demonstrated that in stably stratified stars, random initial magnetic fields can evolve into roughly axisymmetric configurations with both poloidal and toroidal components, which then remain stable over a diffusive timescale (Braithwaite \& Spruit 2004). It would be desirable to provide an analytic proof for the existence of such fields, and to understand how the poloidal and toroidal components help to stabilise each other. Using the energy principle developed by Bernstein et al. (1958), we hope to be able to show that instabilities present for toroidal fields can be eliminated by adding a poloidal component, and vice versa (Akg\"un \& Reisenegger, in prep.) This could also allow us to determine the range of the ratio of the poloidal and toroidal components. Some progress towards that end (using a mixture of analytic and numerical methods) has been made by Braithwaite (2009), where it is found that the toroidal component can be either much stronger than, or of comparable strength to, but {\it not} much weaker than, the poloidal component. Meanwhile, non-axisymmetric equilibria have been found in simulations (Braithwaite 2008). Even in the most strongly magnetised stars, the Lorentz force is still typically a million times weaker than hydrostatic forces due to pressure and gravity. Therefore, a small perturbation in the background equilibrium is sufficient to balance the Lorentz force. In radiative envelopes of massive stars and in degenerate stellar interiors, matter is non-barotropic and stably stratified, which hinders radial displacements --  this allows a wider range of magnetic field structures than found in barotropic fluids (Reisenegger 2009). In Braithwaite et al. (in prep.) we show that a star composed of matter with a barotropic equation of state  is unlikely to retain a fossil field even if equilibria are available, since magnetised regions rise buoyantly and are lost through the surface on a timescale comparable to the reconnection timescale. We also show that analogous equilibria can exist outside of stars, for instance inside radio bubbles inflated by AGN -- see Fig.~\ref{fig:jon}.

It is informative to look at the process of relaxation of an arbitrary field into a stable equilibrium. Equilibria have special topological properities, such as the zero-gravity case where equilibrium requires that $\nabla P = (1/4\pi)(\nabla\!\times\!{\bf B})\!\times\!{\bf B}$ and, since the Lorentz force is perpendicular to the field, the field lines reside in isobaric magnetic surfaces. Relaxation therefore requires topological change, i.e. magnetic reconnection. Magnetic helicity $H\equiv\int {\bf B}\cdot{\bf A}\,{\rm d}V$ where ${\bf B}=\nabla\!\times\!{\bf A}$ is approximately conserved during relaxation, enabling us to predict the energy (and field strength) of the final equilibrium if we know the helicity of the initial turbulent field, since the energy and helicity of the equilibrium are related by the (known) characteristic scale-length of the equilibrium. Unfortunately, we have few constraints on how much helicity a protostellar dynamo is likely to produce (or destroy).

As mentioned above, there are a few mechanisms which cause a fossil field to evolve and which could also bring a core-dynamo field upwards to the surface. First, finite conductivity, which `spreads out' the magnetic energy spatially; the timescale is $\sim10^{10}$yrs in a MS star. Second, buoyancy: imagine a magnetised region surrounded by a less or non-magnetised medium -- since the field provides pressure without mass, the magnetised region must have a lower temperature in order to have the same total pressure and density as its surroundings. Heat diffuses inwards, buoyany rise on a timescale $\tau_{\rm buoy} \sim \beta \,\tau_{\rm K-H}$ where $\beta$ is the ratio of thermal to magnetic pressure (realistically $>10^6$) and $\tau_{\rm K-H}$ is the thermal timescale of the star so that $\tau_{\rm buoy} \succsim10^{12}\,$yrs. In neutron stars there is an equivalent mechanism where the role of temperature is replaced by electron fraction $Y_e$. Finally, meridional circulation: a star in solid-body rotation cannot satisfy both the momentum equation and the heat equation everywhere (the {\it von Zeipel paradox}), and some additional circulation is required (von Zeipel 1924, Zahn 1992, Decressin et al. 2009 and refs. therein.) This circulation takes place on the so-called Eddington-Sweet timescale $\tau_{\rm E-S} \sim \tau_{\rm K-H} (P/P_{\rm breakup})^2$ where $\tau_{\rm K-H}$ is the stellar thermal timescale. This is just possibly relevant for the fastest rotating Ap stars.

In summary, we expect diffusive evolution to be negligible on the MS. In contrast, in neutron stars we expect the diffusive timescales to be rather shorter, in agreement with observational evidence of field decay. Some questions remain: why do not all A stars have these strong fields, and why is there such a large range in field strengths amongst the Ap stars -- at what stage during formation is the Ap destiny of the star determined?

\section{Magnetism and Rotation in the Herbig Ae/Be stars}

Fossil field theory (see \S2) predicts that fields are present throughout the MS and also earlier, in the pre-main sequence (pre-MS) phase. To explain the slow rotation of Ap/Bp stars, magnetic braking during the pre-MS has been proposed.

\begin{figure}[h]
\centering
\includegraphics[width=5.2cm,angle=90]{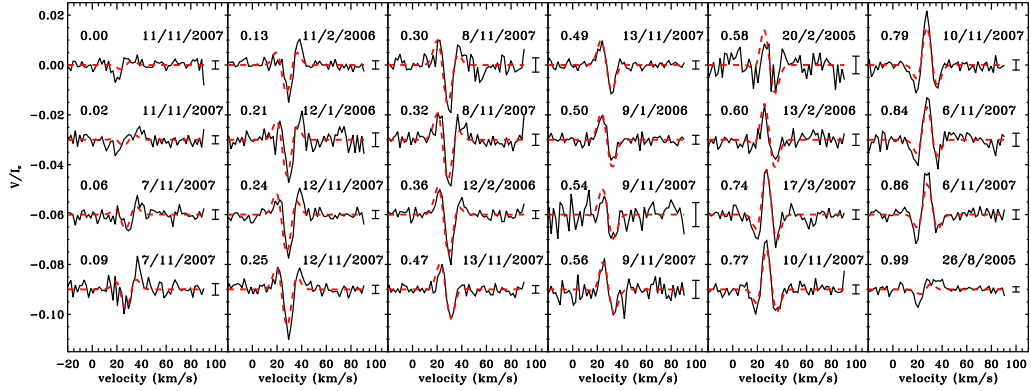}
\caption{LSD $V$ profiles of V380 Ori superimposed by the best oblique rotator model (dashed red line). The numbers close to the profiles are the rotation phase, and the small bars on the right of the profiles are the mean error bars in $V$. The profiles are sorted by increasing rotational phase, and the date of observation is indicated next to each profile (Alecian et al. 2009).}
\label{fig:fitv}
\end{figure}

To test both hypotheses, E.~Alecian and collaborators performed a high-resolution spectropolarimetric survey of over 100 of the pre-MS progenitors of the A and B stars, the Herbig Ae/Be stars, discovering 7 magnetic stars among 128 observed (HD 200775, HD 72106, V380 Ori, HD 190073, NGC 6611 601, NGC 2244 201, NGC 2264 83), implying that $~5$\% of the Herbig Ae/Be stars are magnetic, as predicted by the fossil theory. Furthermore they performed a monitoring of the magnetic stars and were able to fit the temporal variations of their Stokes $V$ profiles using the oblique rotator model (see an example for V380 Ori in Fig. \ref{fig:fitv}). They conclude that these stars host large-scale organised magnetic fields with dipole strengths between 300 G and 3 kG, as predicted by the fossil theory (e.g. Alecian et al. 2008, Alecian et al. 2009, Wade et al., in prep.)

\begin{figure}[h]
\parbox[t]{0.28\hsize}{\caption{\small $v\sin i$ histograms of the magnetic (left) and the non-magnetic (right) HAeBe stars. The y-axes are graduated in percentage of stars on the left, and in number of stars on the right (Alecian et al., in prep.).
\label{fig:statmg}}}\,\,\,\,\,
\parbox[t]{0.70\hsize}{\mbox{}\\
\includegraphics[width=4.8cm,height=4.1cm]{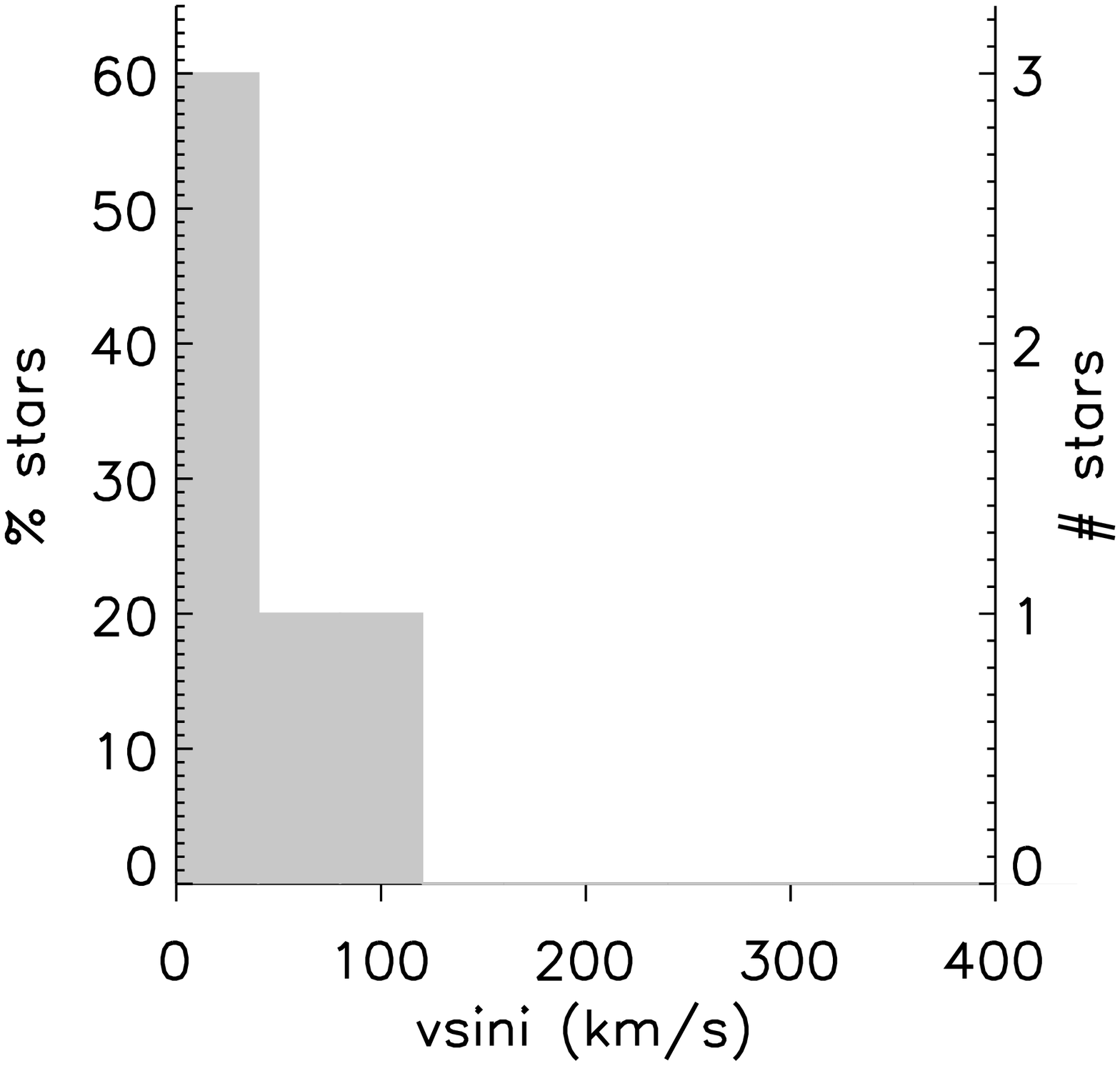}
\includegraphics[width=4.8cm,height=4.1cm]{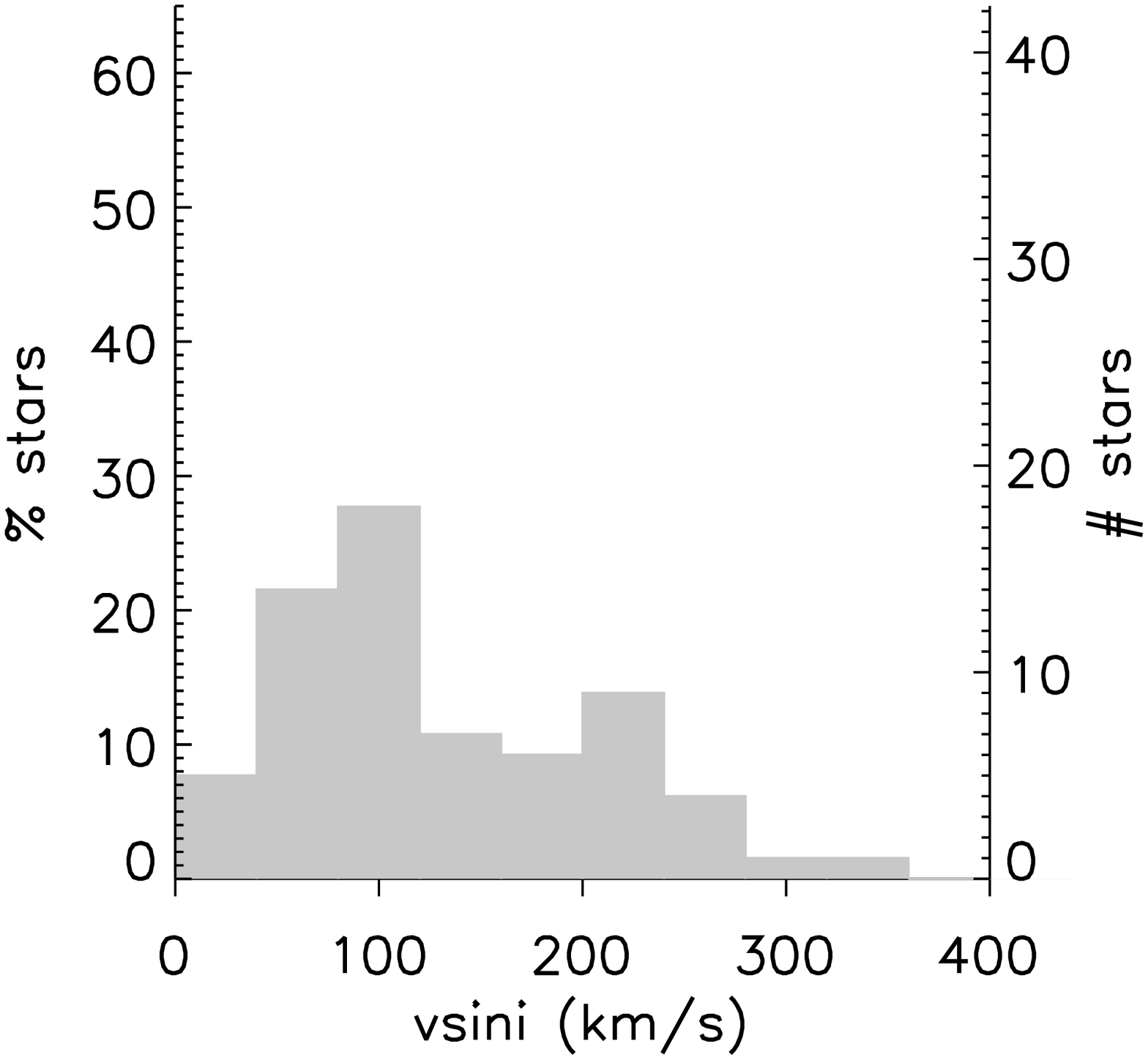}}
\end{figure}

\begin{figure}[h]
\parbox[t]{0.28\hsize}{\caption{\small $v\sin i$ histograms of the normal HAeBe stars projected on the ZAMS (left) and the normal A/B stars (right). The y-axes are graduated in percentage of stars on the left, and in number of stars on the right (Alecian et al., in prep.).
\label{fig:statzams}}}\,\,\,\,\,
\parbox[t]{0.70\hsize}{\mbox{}\\
\includegraphics[width=4.7cm,height=4.2cm]{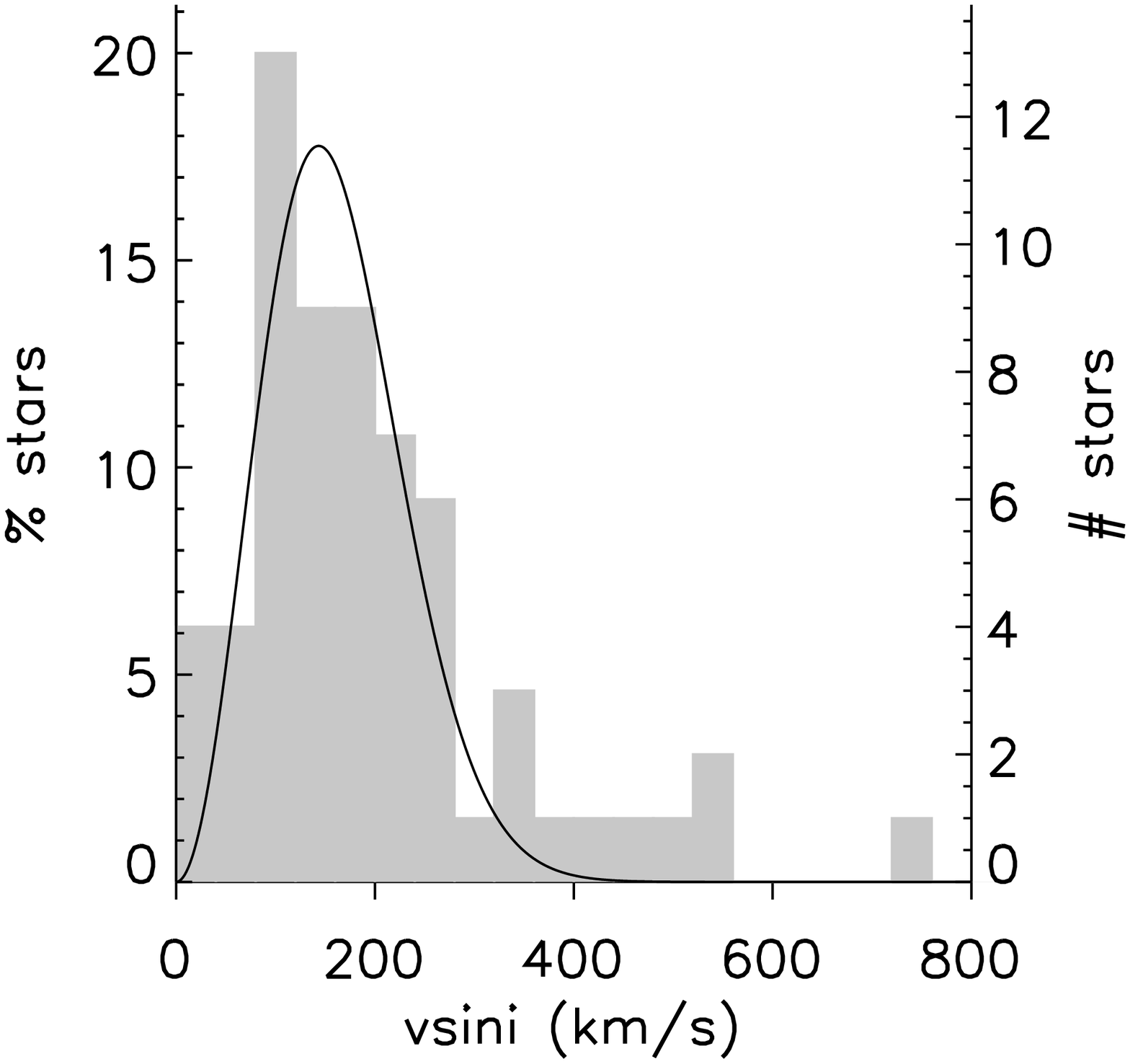}
\includegraphics[width=4.7cm,height=4.2cm]{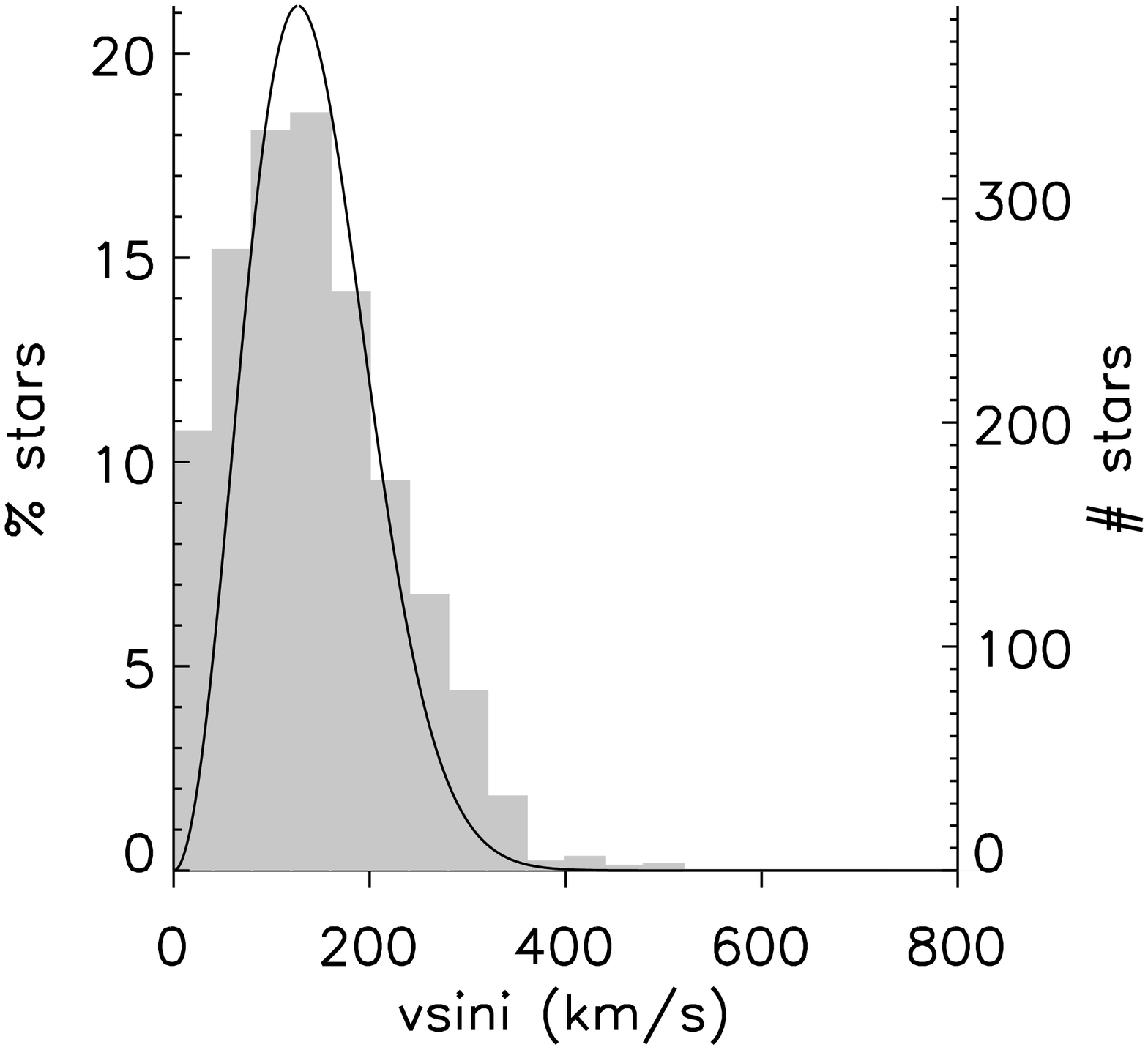}}
\end{figure}

Then they measured the rotation velocities projected on the line of sight ($v\sin i$) of all the stars of their sample, and they first compared the $v\sin i$ distributions of the magnetic to the non-magnetic stars (Fig. \ref{fig:statmg}). They find that the magnetic stars have been braked more than the non-magnetic ones, and that the braking must occur very early during the pre-MS evolution. Finally, they compared the $v\sin i$ distribution of the normal (i.e. non-magnetic and non-binary) HAeBe stars projected onto the zero-age main-sequence (ZAMS) to the normal A/B stars on the MS (Fig. \ref{fig:statzams}). They find that the two distributions are very similar and conclude that the normal HAeBe stars are expected to evolve towards the ZAMS with constant angular momentum (Alecian et al., in prep.)

\vspace{-0.7cm}

\section{Magnetic fields of O and B stars: measurements, statistics and evolution}
\begin{figure}[ht!]
\centering 
 \includegraphics[height=1.5in,width=4.5in,angle=0]{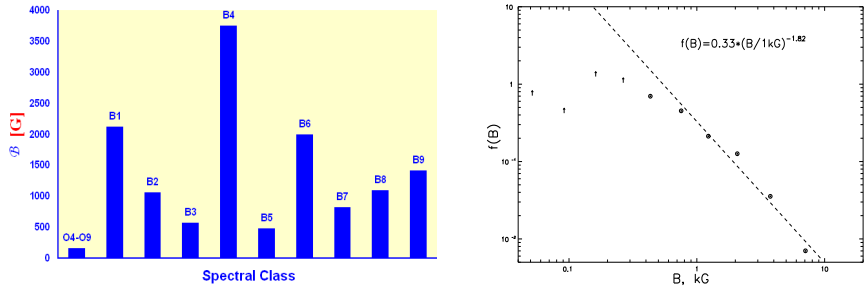}
\caption{\small {\bf Left Panel}: Mean magnetic fields of OB stars for different spectral subclasses.\,\,\,\,\,\,\,
               {\bf Right Panel}: Magnetic field function for B stars.  
        }
\label{Fig.rmsB_MFF}
\end{figure}

Recent measurements of the magnetic fields of early-type stars have been collected in a new catalogue of the magnetic fields of OBA stars~(\cite{Bychkov-2009}). Based on the data from the catalogue together with the newest data, the statistical properties of an ensemble of the magnetic fields of OB stars were investigated. The {\it rms} longitudinal magnetic field $\cal B$ was used as a statistical measure of the mean field. This statistical experiment showed that this value depends weakly on the distribution of the random times of observations and on the structure of the field. The mean magnetic field averaged over the spectral subclasses has an unexpected jump between O and B spectral classes as can be seen in Fig.~\ref{Fig.rmsB_MFF} (left panel). 

We have calculated the differential magnetic field function (MFF) $F({\cal B})$ for B stars, which is defined such that $F({\cal B})d{\cal B}$  is the probability that a {\it rms} longitudinal mean field $\cal B$ lies in the interval $({\cal B}, {\cal B}+ d{\cal B})$ (right panel of Fig.~\ref{Fig.rmsB_MFF}). It is found that the MFF for ${\cal B} > 400\,$G can be approximated by the power function $F(B)= A(B/1kG)^{-\gamma}$ with parameters $A=0.33$ and $\gamma= 1.82$. The MFF steeply decreases for ${\cal B} < 400\,$G -- a result compatible with that of \cite{Auriere-2007} who suggest that weak fields are destroyed by instabilities. It is also found (Kholtygin et al., in prep.) that the {\it rms} mean magnetic field $\cal B$ of the star can decrease by a factor 4-5 times during its evolution from ZAMS to TAMS in accordance with \cite{Landstreet-2008}. 

\section{The effect of atomic diffusion in stellar evolution}
Georges Michaud and his collaborators (O. Richard, J. Richer  and M. Vick) are using abundance anomalies observed in AmFm and Horizontal Branch (HB) stars to constrain the hydrodynamics of stellar interiors.  Their basic code includes all standard equations  of stellar evolution to which they add 56 coupled differential equations to introduce atomic diffusion of all species included in the OPAL Rosseland opacities currently used to model stellar interiors in addition to Lithium, Berillium and Boron, which were calculated by the Montreal group.  They use the original spectra that were used by OPAL to calculate their opacities.  The dominant term in the transport equations involves the difference between gravitational and radiative accelerations.  It is constantly recalculated for every mass shell during the evolution along with the Rosseland opacity so that the calculations take  all  composition changes consistently into account.  For more detail see \cite{RicherMiRoetal98,TurcotteRiMietal98} and \cite{RichardMiRi2001}.

Results indicate that atomic diffusion driven by radiative accelerations can explain the abundance anomalies observed in AmFm and hot HB stars.  In the absence of any competing hydrodynamical process, atomic diffusion actually leads to larger anomalies than observed.  Two competing processes have now been investigated: turbulence mixing a given external mass, and mass-loss at a constant rate.  Observations of Sirius A and of the Hyades star 68 Tau  were found to be explained equally well by either of the two models.  In the case of Sirius,  12 out of the 16 elements have their surface abundances well reproduced either by a mixed outer mass of about $10^{-5}$ solar mass or by a mass-loss rate of $10^{-13}$ solar mass per year.  In the case of 68 Tau, 13 of the 14 elements are obtained within error bars.  The mass-loss rate implies that after $10^8$ years the matter at the surface largely comes from about $10^{-5}$ solar mass below the surface which is also the mixed mass in the model with turbulence.  That is where the important competition between gravitational and radiative accelerations occurs in both models. Observation of  pre-MS  abundance anomalies would favor mass loss as a competing mechanism.  For a detailed description see  \cite{RicherMiTu2000} for the turbulence model and Vick et al.~(2009, in prep.) for the mass loss model.

It is emphasized that there are large differences in the abundances determined by different observers and that a critical analysis of the abundance results is urgently needed to improve the situation and permit more rigorous constraints on the models.

\begin{figure*}[ht!]
\begin{center} 
\includegraphics[scale=.362]{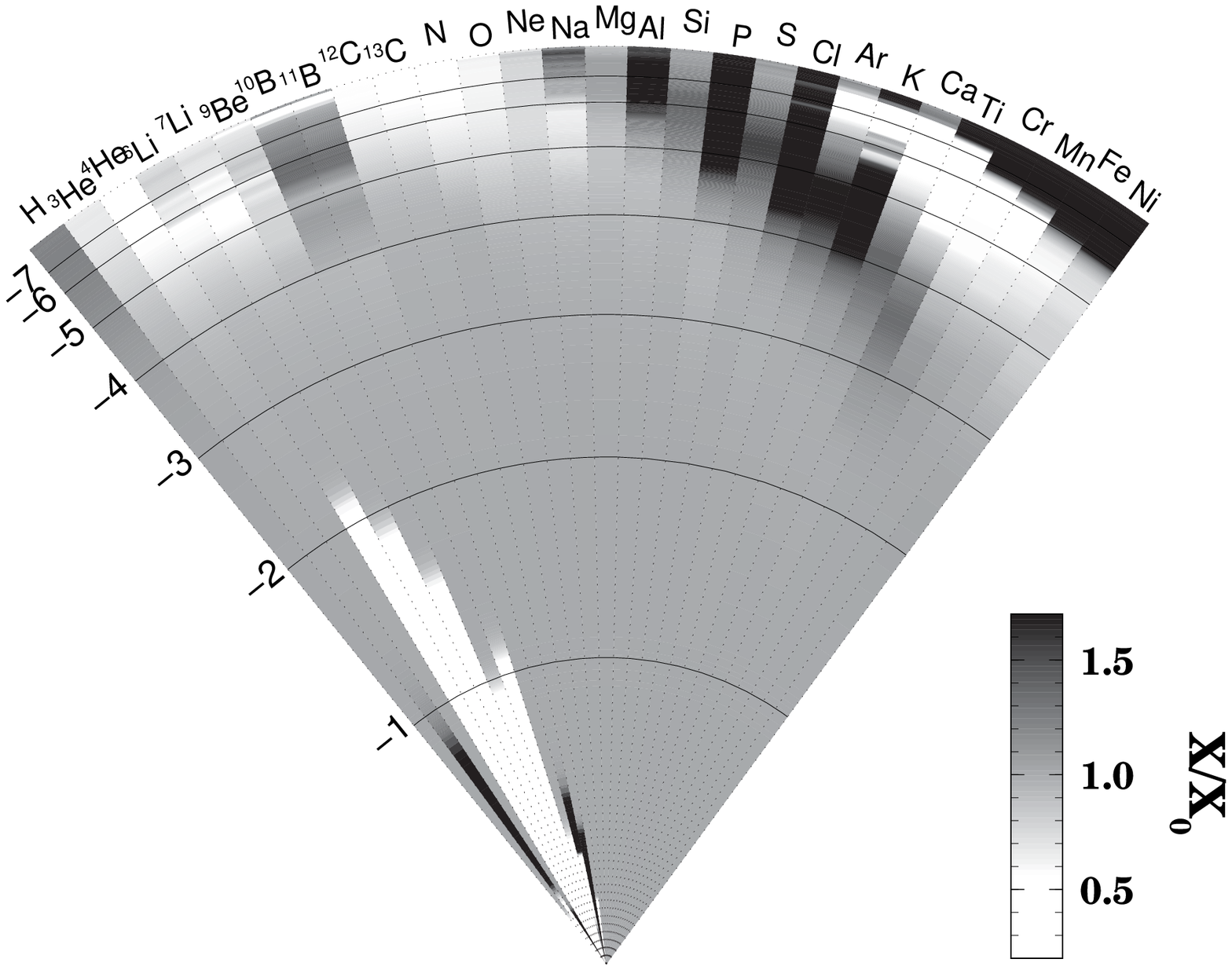}
\includegraphics[scale=.362]{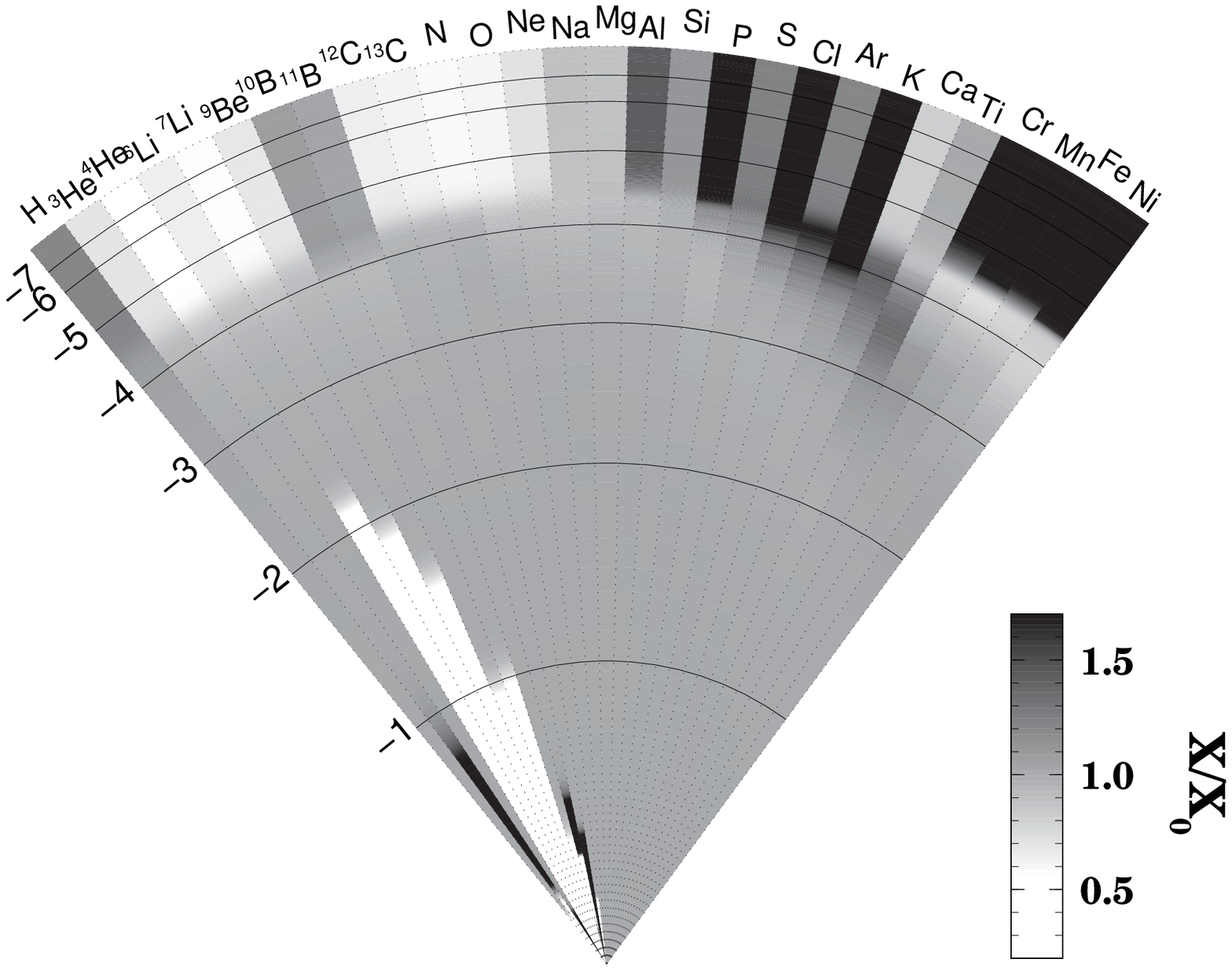}
\caption{\small  Concentration variations for all calculated species in two 2.50 solar mass models at 500 Myr; {\bf Left Panel} with mass loss (10$^{-13} M_\odot/$yr) as the competing process   and  {\bf Right Panel} with turbulence as the competing process. An overabundance by a factor of 1.7 or more appears black in both cases while an underabundance by a factor of 0.5 or less appears white. The radial scale is linear in $r$. Horizontal lines indicate  the mass of the spherical shell  outside a certain radius ($\Delta m$) labeled by $\log (\Delta m/M)$. As is easily seen by comparing the two panels, the interior abundances are quite different in the outer 20\,\% of the radius of the star but, yet, both models have the same surface abundances which also agree reasonnably well with the anomalies observed on some AmFm stars.  See Vick et al.~(2009, in prep.) for details and a similar figure at a different age. The anomalies caused by atomic diffusion affect the outer 30\,\% of the radius while nuclear reactions modify abundances of lighter elements within the inner 50\,\% of the radius leaving only a small buffer where the original abundances of the star have not been modified. 
        }
\label{Fig.fan_michaud}
\end{center}
\end{figure*}

However, even if the surface abundances obtained taking either mass loss or turbulence  into account are very similar, the interior composition is very different -- see Fig.~\ref{Fig.fan_michaud}.  The difference comes from the different nature of the flux in the two models.  In the model with mass loss, the flux is advective and for each species approximately given by $\rho\, c(r,t) (v_D+v_W)$ where $\rho$ is the plasma density, $c(r,t)$ is the local concentration of a given species at time $t$, $v_D$ is the diffusion velocity and $v_W$ the wind velocity.  When the radiative acceleration is much larger than gravity, the $v_D$ is positive and adds up to $v_W$ so that $c(r,t)$ must be smaller for a given mass loss rate.  In those regions of the envelope where radiative acceleration is largest on a given species, the concentration of that species is smallest in a mass loss solution.  Heavy metals then tend to be underabundant between the surface convection zone and $10^{-5}$ solar mass below the surface whereas they are overabundant there in the model with turbulence.  Helium also turns out to be distributed differently in the two models.  This opens the interesting possibility of distinguishing between the two models by asterosismology.  The outer $10^{-5} M_\odot$, although a small mass, occupies approximately 20 \% of the radius of the star and this is where the driving often occurs most efficiently for seismic waves.  The two models should have significantly different asterosismic signatures. These have not yet been calculated in detail.

On the HB, overabundances of Fe by a factor of order 50 are seen in all stars hotter than 11,000\,K in the cluster M15 except for one star which has a relatively  large $v \sin i$.  This can be understood as the consequence of atomic diffusion driven by radiative accelerations as discussed in \cite{MichaudRiRi2008}.  The appearance of abundance anomalies in HB stars, AmFm stars and HgMn stars is explained by their slow rotation if one uses a simple meridional circulation model which involves no adjustable parameters, as is discussed in \cite{QuievyChMietal2009}.  Since this model uses advective flows, it is tempting to assume that the concentration variations seen in the advective flow of the mass loss model would also apply to the meridional flow and this opens the possibility of a test by asterosismology of the nature of the process competing with atomic diffusion as one considers stars with faster rotation in which surface anomalies progressively disappear.

{\bf Diffusion in magnetic stellar atmospheres}
Most magnetic CP stars show abundance variations that are the signature of separation in the atmosphere itself. The modelling of transport processes in the atmosphere is much more demanding than in stellar interiors since the magnetic fields necessitate 3--D modelling for a complete description, which in principle requires 3--D radiative transfer and particle transport.  It should lead to the appearance of abundance spots within the observed magnetic structures.

Results of time dependent diffusion of strontium in magnetic atmospheres were presented by G.~Alecian. In collaboration with M.~Stift he solved in detail the particle transport separately in the presence of horizontal and vertical magnetic fields.  The two solutions are very different. The accumulation can be very short, of the order of days in the high atmosphere. They suggest that as a first approximation, one may consider transport in the vertical direction only, since the distances are shorter; the spots are then created by the effect of the magnetic field on the vertical diffusion velocity. These results are very encouraging but many steps are still required to obtain a map of abundance anomalies at the surface of magnetic CP stars.

On the other hand, it was suggested by J.~Portnoy and R.~Steinitz using qualitative arguments that magnetic field gradients could lead to an additional term in the particle transport equation, perhaps explaining the over-abundance of heavier elements,
 and that dielectronic recombinations could be potentially important for the ionization equilibria in magnetic atmospheres.


\section{Towards a coherent picture of internal transport 
in CP stars}

The study of asteroseismology and powerful ground-based instrumentation dedicated to stellar physics is developing strongly (CoRoT, KEPLER, ESPaDOnS, etc.) generating tight constraints on the internal structure, dynamics, and magnetism of stars. For this reason stellar models are needed that include transport processes from the birth of stars to their death. A coherent picture of the dynamics of stellar radiative zones, where non-standard chemical mixing occurs, is thus required (cf. Zahn 1992, Meynet \& Maeder 2000). A complex transport, {\it the rotational transport}, which involves several mechanisms, takes place in these regions.

First, rotation, structural adjustments and angular momentum losses at the surface induce `baroclinic' or `meridional circulation' (see \S\ref{sec:equilibria}; Maeder \& Zahn 1998, Mathis \& Zahn 2004, 2005) which acts simultaneously to transport angular momentum, chemicals and magnetic field by advection as well as inducing differential rotation. Next, differential rotation induces hydrodynamical turbulence via shear, baroclinic, multidiffusive and other instabilities, which of course in turn damps the differential rotation -- as happens in the terrestrial atmosphere (Zahn 1983; Mathis et al. 2004 and refs. therein). Then, (differential) rotation interacts with fossil magnetic fields. In this case, the mean secular torque of the Lorentz force (Gough \& McIntyre 1998, Mathis \& Zahn 2005,  Brun \& Zahn 2006) and magnetohydrodynamical instabilities such as Tayler and multidiffusive magnetic instabilities (Tayler 1973, Spruit 1999, Menou et al. 2004) can modify the transport of angular momentum and chemicals. Eventually, a dynamo in stably-stratified stellar radiation zone is possible due to the non-linear interaction of instability-velocity field and instability-magnetic field but its detection in numerical simulations is still under exploration (Spruit 2002, Braithwaite 2006, Zahn et al. 2007).

Finally, internal gravity waves are excited at the boundaries with convective zones. They propagate through radiative regions where they extract or deposit angular momentum at the location where they are damped, leading to a modification of the angular velocity profile and consequently of the chemical distribution (Goldreich \& Nicholson 1989; Talon \& Charbonnel 2005). Because of their frequencies, they can be strongly modified by both the Coriolis acceleration (Mathis et al. 2008, Mathis 2009) and the Lorentz force (Kumar et al. 1999).

If the star is one component of a close binary system, tidal processes modify the angular momentum transport via the external torque of the equilibrium tide and the dynamical tide, i.e. the internal gravity waves which are excited by the tidal potential (Zahn 1977). Like external winds, tides can strongly modify the internal angular velocity distribution and thus the magnetic behaviour and mixing.

\begin{figure}[h]
\parbox[t]{0.24\hsize}{\caption{The complete rotational transport in a stellar radiation zone.
\label{fig:mathis}}}\hspace{-0.8cm}\vspace{-1.1cm}
\parbox[t]{0.77\hsize}{\mbox{}\\ \includegraphics[width=0.75\hsize,angle=90]{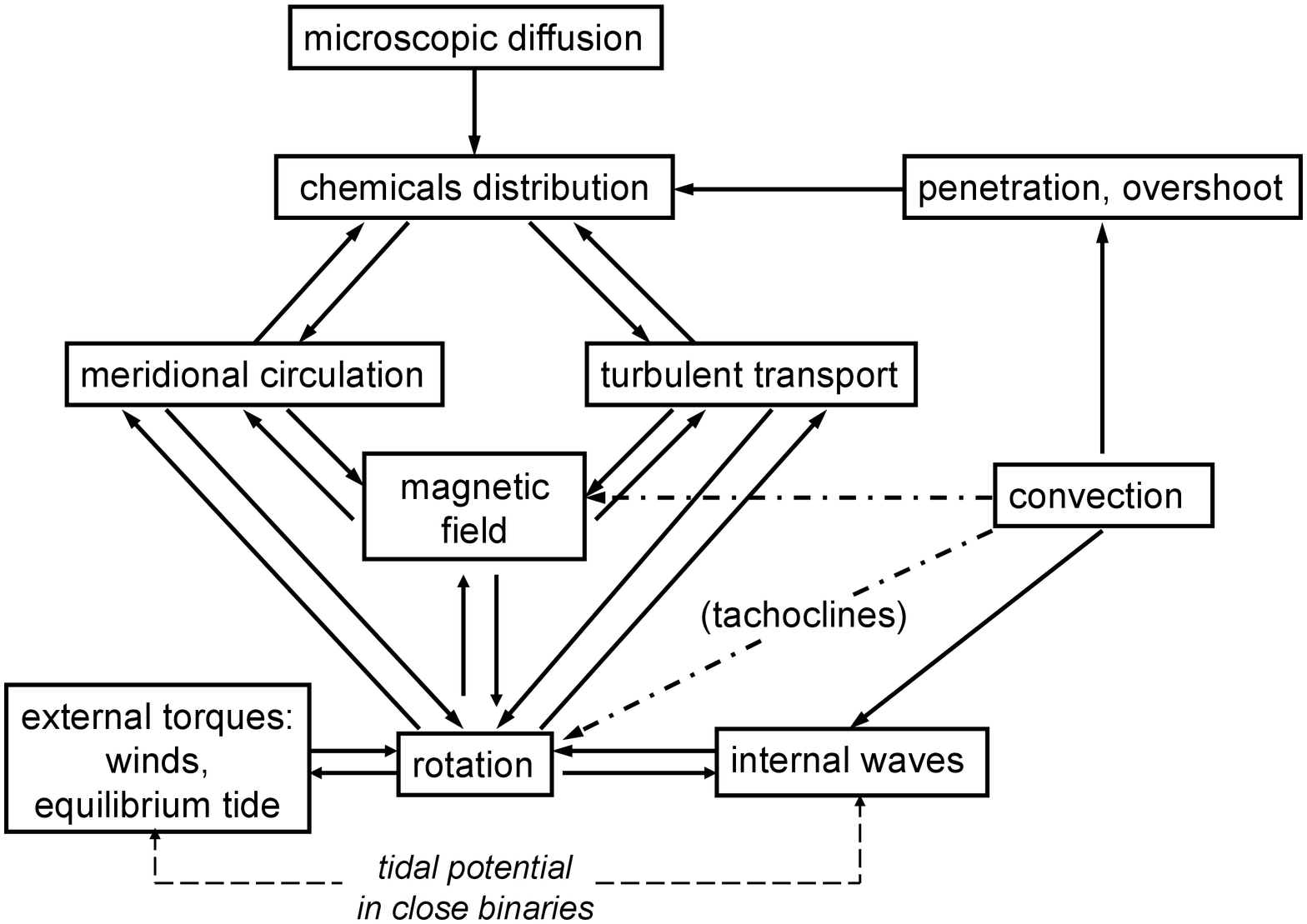}}
\end{figure}

The rotational transport and various interactions are summarized in Fig.~\ref{fig:mathis}. These mechanisms now have to be applied to massive peculiar stars, to study their rotational history and their potential equilibrium state such as the magnetohydrostatic equilibrium (see Moss 1977, Duez et al. 2009, and \S2).

\begin{thebibliography}{}

\bibitem[]{alecian08}
     {Alecian, E., Catala, C., Wade, G.A. et al.} 2008,
     \textit{MNRAS} 385, 391
\bibitem[]{alecian09}
     {Alecian, E., Wade, G.A., Catala, C. et al.} 2009,
     \textit{MNRAS}, in press (arXiv:0907.5113)
\bibitem[Auri\`ere et al. (2007)] 
   {Auriere-2007}
 {Auri\`ere, M., Wade, G.A., Silvester, J. et al.}, 2007, \textit{A\&A}, 475, 1053
\bibitem[Babel \& Montmerle (1997)]{Bab_and_Mon_1997}Babel, J. \& Montmerle, T. 1997, A\&A 323, 121
\bibitem[Bernstein et al. (1958)]{Bernstein58}
     {Bernstein, I.B., Frieman E.A., Kruskal M.D. \& Kulsrud R. M.} 1958, \textit{Proc. R. Soc. A} 244, 17
\bibitem[Bouret et al. (2008)]{Bouret_et_al_2008}Bouret, J.-C., Donati, J.-F., Martins, F. et al. 2008, MNRAS 389, 75
\bibitem[]{} Braithwaite, J. 2006, \textit{A\&A} 449, 451
\bibitem[Braithwaite (2008)]{Braithwaite08}
     {Braithwaite}, J. 2008, \textit{MNRAS} 386, 1947
\bibitem[Braithwaite (2009)]{Braithwaite09}
     {Braithwaite}, J. 2009, \textit{MNRAS} 397, 763
\bibitem[Braithwaite \& Spruit (2004)]{Braithwaite04}
     {Braithwaite, J. \& Spruit, H.} 2004, \textit{Nature} 431, 819
\bibitem[]{} Brun, A. S. \& Zahn, J.-P. 2006, \textit{A\&A} 457, 665
\bibitem[Bychkov et al. 2009] 
   {Bychkov-2009}
 {Bychkov, V.D., Bychkova, L.V. \&  Madej, J.}, 2009, \textit{MNRAS}, 394, 1338
\bibitem[]{} Decressin, T., Mathis, S., Palacios, A., Siess, L., Talon, S., Charbonnel, C. \& Zahn, J.-P. 2009, \textit{A\&A} 495, 271
\bibitem[Donati \& Landstreet (2009)]{Don-Lan-09}
      {Donati, J.-F. \& Landstreet, J.D.} 2009, ARAA 47,333.
\bibitem[]{} Duez, V., Mathis, S. \& Turck-Chi\`eze, S. 2009, \textit{MNRAS}, accepted
\bibitem[]{} Goldreich, P. \& Nicholson, P. D. 1989, \textit{ApJ} 342, 1079
\bibitem[Goossens et al. (1981)]{Goossens81}
     {Goossens, M., Biront, D. \& Tayler, R.J.} 1981, \textit{Ap\&SS} 75, 521
\bibitem[]{} Gough, D. O. \& McIntyre, M. E. 1998, \textit{Nature} 394, 567
\bibitem[]{} Kumar, P., Talon, S. \& Zahn, J.-P. 1999, \textit{ApJ} 520, 859 
\bibitem[Landstreet \& Borra (1978)]{Lan_and_Bor_1978}Landstreet, J.D. \& Borra, E.F. 1978, ApJ 224, L5
\bibitem[Landstreet et al. (2008)]{Landstreet_et_al_2008}Landstreet, J.D., Silaj, J., Andretta V. et al. 2008 A\&A, 481, 465
\bibitem[]{} Maeder, A. \& Zahn, J.-P. 1998, \textit{A\&A} 334, 1000
\bibitem[Markey \& Tayler (1973)]{Markey73}
     {Markey, P. \& Tayler, R.J.} 1973, \textit{MNRAS} 163, 77
\bibitem[]{} Mathis, S. 2009, \textit{A\&A}, 506, 811
\bibitem[]{} Mathis, S., Palacios, A. \& Zahn, J.-P. 2004, \textit{A\&A} 425, 243
\bibitem[]{} Mathis, S. \& Zahn, J.-P. 2004, \textit{A\&A} 425, 229
\bibitem[]{} Mathis, S. \& Zahn, J.-P. 2005, \textit{A\&A} 440, 653
\bibitem[]{} Menou, K., Balbus, S. A. \& Spruit, H. C. 2004, \textit{ApJ} 607, 564
\bibitem[]{} Meynet, G. \& Maeder, A. 2000, \textit{A\&A} 361, 101
\bibitem[{{Michaud} {et~al.} (2008)}]{MichaudRiRi2008}
{Michaud}, G., {Richer}, J. \& {Richard}, O. 2008, ApJ, 675, 1223
\bibitem[]{} Moss, D. 1977, \textit{MNRAS} 178, 51
\bibitem[{{Quievy} {et~al.}(2009)}]{QuievyChMietal2009}
{Quievy}, D., {Charbonneau}, P., {Michaud}, G. \& {Richer}, J. 2009, A\&A,
  500, 1163
\bibitem[Reisenegger (2009)]{Reisenegger09}
     {Reisenegger, A.} 2009, \textit{A\&A} 499, 557
\bibitem[{Richard {et~al.} (2001)}]{RichardMiRi2001}
Richard, O., Michaud, G., \& Richer, J. 2001, ApJ, 558, 377
\bibitem[{Richer {et~al.} (1998)}]{RicherMiRoetal98}
Richer, J., Michaud, G., Rogers, F., {et~al.} 1998, ApJ, 492, 833
\bibitem[{Richer {et~al.} (2000)}]{RicherMiTu2000}
Richer, J., Michaud, G., \& Turcotte, S.: 2000, ApJ, 529, 338
\bibitem[]{} Spruit, H. C. 1999, \textit{A\&A} 349, 189
\bibitem[]{} Spruit, H. C. 2002, \textit{A\&A} 381, 923
\bibitem[]{} Talon, S. \& Charbonnel, C. 2005, \textit{A\&A} 440, 981
\bibitem[Tayler (1973)]{Tayler73}
     {Tayler, R.J.} 1973, \textit{MNRAS} 161, 365
\bibitem[Townsend et al. (2007)]{Townsend_et_al_2007}Townsend, R.H.D., Owocki, S.P. \& Ud Doula, A. 2007, MNRAS 382, 139; see also http://www.astro.wisc.edu/$\sim$townsend/static.php?ref=rrm-movies
\bibitem[{Turcotte {et~al.}(1998)}]{TurcotteRiMietal98}
Turcotte, S., Richer, J., Michaud, G., Iglesias, C., \& Rogers, F. 1998, ApJ,
  504, 539
\bibitem[Wright (1973)]{Wright73}
     {Wright, G.A.E.} 1973, \textit{MNRAS} 162, 339
\bibitem[]{} Zahn, J.-P. 1977, \textit{A\&A} 57, 383
\bibitem[]{} Zahn, J.-P. 1983, \textit{Saas-Fee Advanced Course 13, Astrophysical Processes in Upper Main Sequence Stars, eds. B. Hauck and A. Maeder, publisher: Geneva Observatory} 253
\bibitem[]{} Zahn, J.-P. 1992, \textit{A\&A} 265, 115
\bibitem[]{} Zahn, J.-P., Brun, A. S. \& Mathis, S. 2007, \textit{A\&A} 474, 145
\bibitem[von Zeipel (1924)]{vonZeipel_1924}von Zeipel, H. 1924, MNRAS 84, 665
\end{thebibliography}
\end{document}